\documentclass[12pt]{article}

\usepackage{amsfonts}
\usepackage{amsmath}
\usepackage[includeheadfoot,left=1.in,right=1.in,top=1.in,bottom=1.in,bindingoffset=0.in,nohead]{geometry}
\usepackage{graphicx}
\usepackage{graphics}
\usepackage{subfigure} 
\usepackage{booktabs}
\usepackage{natbib}
\usepackage[figuresright]{rotating}
\usepackage{graphics}
\usepackage[onehalfspacing]{setspace}
\usepackage{appendix}
\usepackage{caption}
\usepackage[pdfborder={0 0 0}]{hyperref}
\usepackage[usenames,dvipsnames]{xcolor}

\definecolor{mygreen}{rgb}{1.0000,    0.4,         0}

\begin{document}

\title{On the Past, Present, and Future\\of the Diebold-Yilmaz Approach to \\Dynamic Network Connectedness}

\author{Francis X. Diebold  \\ University of Pennsylvania \and Kamil Yilmaz \\  Ko\c{c} University  \\ \bigskip
}

\date{\today }
\maketitle
\thispagestyle{empty}

\bigskip

\bigskip

\begin{spacing}{1}
\noindent {Abstract}:  We offer retrospective and prospective assessments of the Diebold-Yilmaz connectedness research program, combined with personal recollections of its development.  Its centerpiece in many respects is \cite{DieboldandYilmaz2014}, around which our discussion is organized.
 
\end{spacing}

\bigskip

\bigskip

\bigskip

\begin{spacing}{1}

\noindent {Acknowledgments}:  We are grateful to the editors for selecting our paper, and to our many colleagues and students --  in particular our wonderfully-connected network of ``connectedness coauthors" --  for their myriad contributions to the research program described here and their friendship over the years.  We are also grateful to Tim Bollerslev, who provided  helpful comments on an earlier draft. Yilmaz gratefully acknowledges the support of The Scientific and Technological Research Council of Turkey (TUBITAK) under grant number 121C271.  All remaining errors of omission or commission are our alone.

\end{spacing}

\bigskip

\bigskip

\bigskip

\noindent {Key Words}: Contagion, spillovers, financial markets, vector autoregressions, variance decompositions.

\bigskip

\noindent {JEL codes}: C1, C3, G1

\bigskip

\noindent Corresponding author:  F.X. Diebold, fdiebold@sas.upenn.edu

\newpage

%
%
%

\setcounter{page}{1}
\thispagestyle{empty}

It is a pleasure to help celebrate the fiftieth birthday of the \textit{Journal of Econometrics}, and a great honor to have  \cite{DieboldandYilmaz2014}  selected as one of the five papers included in this 50th Jubilee Issue.    In this note we offer retrospective and prospective assessments of the Diebold-Yilmaz connectedness research program, combined with personal recollections of its development.  

\section{Origins in the Asian Contagion}
\label{intro}

It all began more than thirty years ago, when FD was an Economist at the Federal Reserve Board in Washington, DC, and KY was a Ph.D. student at the University of Maryland.  In the late 1980s John Haltiwanger had kindly invited FD to teach an advanced  Ph.D. macro-econometrics course at Maryland, and KY was a student in the course.  We had many stimulating discussions, some of which eventually produced, for example, \cite{DGY1994}. But our paths diverged when FD went to the University of Pennsylvania as an assistant professor in 1989, and KY went to Ko\c{c} University as an assistant professor in 1994 after working for two years as an Economist at the World Bank.  The story might naturally have ended there, but no.  

There was indeed a  long gap, but then KY took a sabbatical at Penn during the 2003-2004 academic year.  There was no real goal other than general intellectual stimulation for each of us.  In particular, we had no joint work in progress, and indeed no real plans for joint work,  when KY and his family arrived in Philadelphia in the summer of 2003.  But of course we subsequently had many discussions on  various issues and ideas.  

One thing that intrigued us both was financial market contagion.  In particular, the infamous 1997 sequential collapse of several far-eastern currencies -- the ``Asian Contagion" -- was still fresh in the professional  consciousness, and hence in ours.  How to define ``contagion"?  How to measure it?  Did it really exist?  We had many conversations, typically on the ninth floor of the Wharton School's Huntsman Hall, after visiting the adjacent coffee room, looking out over the Philadelphia skyline.

\section{Variance Decompositions and Network Connectedness}

The Diebold-Yilmaz dynamic connectedness measurement framework grew from those Huntsman Hall conversations. It seemed natural to define \textit{h}-step connectedness in terms of variance decompositions. We liked the fact that  variance decompositions naturally promote uniform measurement across variables via their ``fraction of h-step-ahead forecast error variance" perspective, whereas other possible approaches based for example on impulse-response functions did not, so we simply proposed the variance decomposition approach, directly.

  That is, we proposed measuring what we would later call ``\textit{h}-step pairwise directional connectedness from \textit{j} to \textit{i}" as the fraction of \textit{h}-step forecast error variance of variable \textit{i} arising from shocks to variable \textit{j}.  Allowing for time-varying parameters in the model from which the variance decompositions were calculated, moreover, allowed for the important  possibility of time-varying connectedness.

  Our goal was always the empirical  description of connectedness and its evolution, ``getting the facts straight" with minimal assumptions.  In particular, we didn't want to have to take a stand on  deep underlying structural mechanics, and indeed our intentionally reduced-form measurements are consistent with a wide variety of possible underlying structures.  This is important in economic and financial environments, where connectedness and its evolution are typically incompletely understood at best.

Initially we didn't use the term  ``connectedness". Instead we used more traditional and perhaps more exciting, if nevertheless only vaguely-defined, terms like ``contagion" or ``spillovers".  It soon became clear, however, that different research tribes had  different,  strongly-held, and sometimes  conflicting views about the meanings of such terms.  We wanted an uncontroversial term that simultaneously conveyed the essence of our measure, and we eventually settled on ``connectedness".

\cite{DieboldandYilmaz2009} contains the  embryonic idea. We obtain the variance decomposition using a VAR approximating model, featuring  Cholesky factor identification,  small data, and no awareness of network theory or graphics.\footnote{For some colorful background  on  ``small data" vs. ``Big Data" definitions and history, see \cite{Diebold2021podcast}.} The empirical work focuses on volatility connectedness in equity markets, although the term ``connectedness" is  not used.

  
 \cite{DieboldandYilmaz2012} takes  a significant step forward, moving to the generalized identification of \cite{PesaranShin98}, which builds on  \cite{KoopPesaranPotter96}.  In the generalized identification environment, unlike with Cholesky factor identification,  there is no issue of variable  ordering.  Equally importantly,  \cite{DieboldandYilmaz2012} also  considers not only ``total connectedness" but also  ``directional" aspects, so that $i {\rightarrow} j$ connectedness need not match $j {\rightarrow} i$ connectedness.  The term ``connectedness" is still never used, however, and network perspectives are absent.  The empirical work again involves only small data, focusing on four daily U.S. asset classes (stocks, bonds, commodities, and foreign exchange).

 \cite{DieboldandYilmaz2014}, included in this 50th Jubilee Issue of \textit{Journal of Econometrics}, is the real breakthrough. We of course maintain  generalized identification  and directional perspective, but crucially, we also couch the analysis in a network framework for the first time, and we introduce some  crude network graphics.  We also use the ``connectedness" language throughout.
 
 \begin{table}[t] \large 
 	\caption{\large Table 1: Connectedness Table /  Network Adjacency Matrix, $D$} 
 	  \vspace*{-4mm}
 	 	\begin{center}
 		\begin{tabular}{lccccc}
 			\toprule
 			& \bf $x_1$ & \bf $x_2$ & \bf $...$ & \bf $x_N$ &   From Others to $i$ \\
 			&&&&& (In-Degrees) \\ \midrule
 			\bf $x_1$ & $ d_{11}^H$ & $d_{12}^H$ & $\cdots$ &  $d_{1N}^H$ & $\Sigma_{j=2}^N d_{1j}^H$ \\
 			\bf $x_2$ & $d_{21}^H$ & $d_{22}^H$ & $\cdots$ & $d_{2N}^H$ & $\Sigma_{j=1, j \ne 2}^N d_{2j}^H$  \\
 			\bf $\vdots$ & $\vdots$& $\vdots$ & $\ddots$ & $\vdots$ & \vdots \\
 			\bf $x_N$ & $d_{N1}^H$ & $d_{N2}^H$ & $\cdots$ & $ d_{NN}^H$ & $\Sigma_{j=1}^{N-1} d_{Nj}^H$  \\ \midrule
 			To Others From $j$      &   {$\Sigma_{i=2}^N d_{i1}^H$}  & {$\Sigma_{i=1, i \ne 2}^N d_{i2}^H$}     & {$\cdots$}    & {$\Sigma_{i=1}^{N-1} d_{iN}^H$}    &  {$\Sigma_{i,j=1; i \ne j}^N d_{ij}^H$}  \\  
 			(Out-Degrees)    &     &      & {}    &     &  \\ 
 			\bottomrule
 		\end{tabular}
 	\end{center} \normalsize
 	Notes: We show an illustrative connectedness table (weighted, directed network adjacency matrix) based on an $N$-variable $H$-step variance decomposition, $D$.  The fraction of variable $i$'s $H$-step-ahead forecast error variance due to shocks in variable $j$ is $d^H_{ij}$.  Equivalently, $H$-step pairwise directional connectedness from $j$ to $i$, $ C_{i \leftarrow j}^H$, is $d^H_{ij}$.  Total directional connectedness from others to $i$ (the in-degree, or from-degree of network node $i$) is given by  $ C_{i \leftarrow \bullet}^H = \sum_{ j\ne i} d_{ij}^H$, and  total directional connectedness to others from $j$ (the out-degree, or to-degree of network node $j$) is  $ C_{\bullet \leftarrow j}^H =   \sum_{ i\ne j} d_{ij}^H$.  Total system-wide connectedness is the grand sum of all off-diagonal elements, $C^H= \sum_{i, j; i \ne j} d^H_{ij}$. 
 		\label{template} 
 \end{table}

 Let us provide some background  on  \cite{DieboldandYilmaz2014}.  Around the time when the paper was written, it was hard to go through a week in a department of economics, statistics, computer science, etc., without encountering numerous seminars and conversations about networks -- network formation, network structure, etc. The excitement was palpable.  Surely, we thought, our work on  connectedness (as we had by then begun to call it) must be somehow related to the characterization of network structure.  Unfortunately, however,  we  knew very little about the characterization of network structure! 
 
 So we took the plunge, starting with  \cite{Newman2010} and then working forward and backward, and our eyes were opened.  What a playground:  weighted, directed networks and their node degrees; in-degrees and out-degrees; degree distributions; numerous so-called ``centrality" measures; and on and on.   The key insight -- the real ah-ha! moment -- came when we realized that the variance decomposition matrix, or ``connectedness table", at the center of the Diebold-Yilmaz research program could be viewed as the  adjacency matrix of a weighted, directed network, as illustrated in Table \ref{template}.   Hence the powerful toolkit of methods for network measurement and characterization immediately applies to connectedness.  Indeed several of the connectedness measures that we had independently proposed (total directional connectedness to and from) are prominent network statistics (out-degrees and in-degrees).

Now let us move ahead a few more years to  \cite{Demireretal2018}, who take the connectedness framework closer to completion, moving to Big Data environments with regularized estimation and sophisticated  network graphics that scale effortlessly with network dimension. First consider  regularized estimation.  In contrast to a typical small-data VAR circa 1985 (perhaps a 5-variable VAR(3), which can be estimated by OLS), regularization is \textit{essential} in Big Data environments circa 2023 (perhaps a 500-variable VAR(3), to be used for connectedness measurement).\footnote{Interestingly, even in small-data VAR environments some regularization in the form of Bayesian shrinkage toward random walk dynamics has long been recognized as helpful; see \cite{Doanetal1984}.}   \cite{Demireretal2018} use the LASSO \citep{tibshirani1996regression}, which selects and shrinks.

Second,  consider  network graphics.  If regularization is the key to producing credible network VAR estimates in high dimensions, then graphics are the key to  {understanding}  those estimates.  To see why, consider first the  small-data 5-variable VAR(3) case.  We can't stare productively at nearly 100 estimated  coefficients and hope to learn much, but we can certainly stare productively at a 5$\times$5 connectedness table, as per Table \ref{template}. That is, variance decompositions save the day, rendering dozens of uninterpretable estimated coefficients economically interpretable by viewing them through a particular lens,  which is a key part of the message of \cite{Sims1980}.  And crucially,  \textit{tabular presentation is adequate}.   In contrast, a  Big Data 500-variable VAR(3) environment causes inescapable trouble.  Of course we can't stare productively at VAR coefficients -- now in the  hundreds of thousands -- but we also can't stare productively at a 500x500 connectedness table!  Tabular presentations of variance decompositions, which save the day in small-data environments, are themselves unworkable in high dimensions.

\begin{figure}[t]
	\begin{center}
		\caption{\large Figure 1: Illustrative Network ``Spring Graph"}
		\label{fig:spring}
		\includegraphics[trim = 0mm 0mm 0mm 0mm, clip, scale=.35]{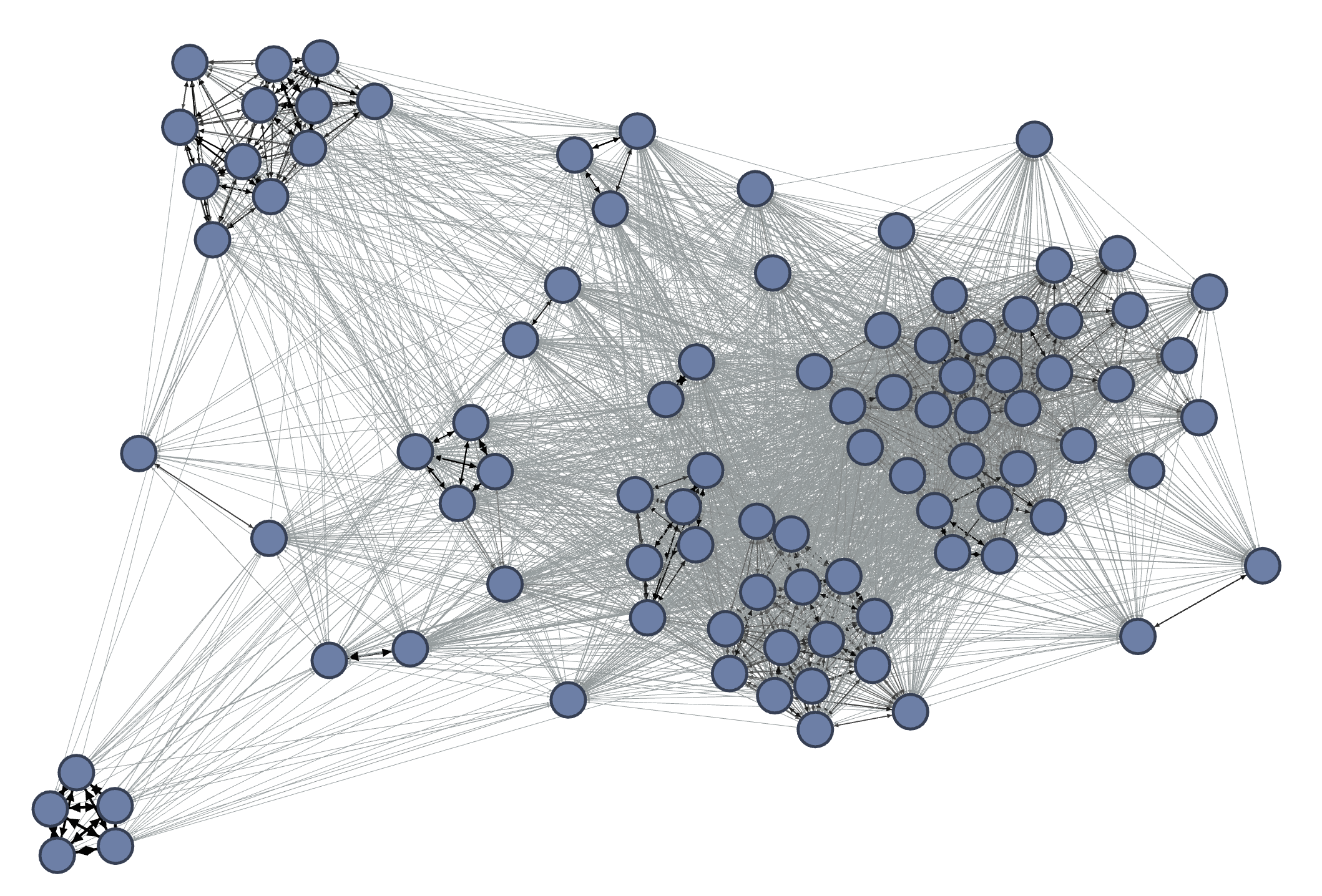}
	\end{center}
	Notes: The underlying  algorithm \citep{jacomy2014forceatlas2} finds a steady state in which repelling and attracting forces exactly balance, where nodes repel each other, and  links, like springs, attract their nodes with force proportional to average pairwise directional connectedness ``to" and ``from."
\end{figure}

But network tools again come to the rescue, this time by providing  highly-revealing and readily-scalable visualizations of variance decompositions, as in the illustrative  ``spring graph" shown in Figure \ref{fig:spring}, which is immediately interpretable, whether for a network of dimension 5 or 500 or 5000.  In particular, network graphics are far from pretty afterthoughts; rather, they are central to the connectedness framework, facilitating the \textit{understanding} of estimated high-dimensional approximating models by \textit{visualizing} their variance decompositions. Even without knowing the details of how Figure \ref{fig:spring} is constructed, for example, a viewer can immediately see two small and very tightly-connected clusters in the upper-left and lower-left, and two large and somewhat less-tightly connected clusters in the upper- and lower-right.\footnote{For details of spring graph construction, see the notes to  Figure \ref{fig:spring}.}$^,$\footnote{For readers curious as to the actual network represented by  Figure \ref{fig:spring}, see  \cite{Demireretal2018}.}

Finally, as mentioned earlier,  an important part of the Diebold-Yilmaz framework is allowing for time-varying coefficients  in approximating models (whether by rolling the estimation or by   explicitly modeling time-varying coefficients), and hence allowing for time-varying variance decompositions that translate into time-varying network graphs.  There will then be a \textit{different} Table \ref{template} and Figure \ref{fig:spring} in each period as the approximating model coefficients  evolve, and one could  watch an animation of Figure \ref{fig:spring}  over time.\footnote{Some details remain to be worked out, such as how to ``anchor" the set of graphs, and we look forward to someone doing so.}  As a practical matter, however, we are often content to examine simple time-series plots of various elements of a time-varying connectedness table  obtained from a rolling estimation window. 

\section{Parallels in the Measurement of International Trade and Systemic Financial Risk}

We have already discussed the deep connections between our connectedness measures and traditional tools for characterizing network structure.  Here we  briefly discuss some other deep connections, first to the measurement of international trade and then to the measurement of systemic financial risk.

\subsection{International Trade}

An ``import/export" perspective provides immediate insight on our connectedness measures, at different levels of aggregation from the most granular to the most aggregative. In an obvious notation, pairwise directional connectedness  $ C_{i \leftarrow j}^H=d^H_{ij} $ is ``$i$'s imports from $j$".  On net,   $C_{ij}^H=C_{j \leftarrow i}^H-C_{i \leftarrow j}^H$ is the ``$ij$ bilateral trade balance".   Total directional connectedness from others to $i$,  $ C_{i \leftarrow \bullet}^H = \sum_{ j\ne i} d_{ij}^H$ is ``$i$'s total imports", and  total directional connectedness to others from $j$,   $ C_{\bullet \leftarrow j}^H =   \sum_{ i\ne j} d_{ij}^H$ is ``$j$'s total exports".   On net, $ C_i^H =  C_{\bullet  \leftarrow i}^H - C_{i \leftarrow \bullet}^H $ is ``$i$'s multilateral trade balance".  System-wide connectedness,  $  C^H= \sum_{i \ne j} d^H_{ij}$ is ``total world exports" (or imports, since they must agree at the world level).

\subsection{Systemic Financial Risk}

Marginal expected shortfall, $MES$, is an important measure of systemic financial risk; see \cite{Acharyaetal2012} and the references therein. It is essentially a measure of the sensitivity of an individual firm's return to extreme market-wide returns.  In particular, 
$$
MES^{j|mkt}=E ( r_{j} \, | \, 
\mathbb{C}(r_{mkt}) ),
$$
where $r_j$ denotes firm $j$'s return and  $\mathbb{C}(r_{mkt})$ an extreme market return event.  $MES$ is effectively a market-based ``stress test" of firm $j$'s fragility during systemic events.  It is also clearly and intimately related to  Diebold-Yilmaz  ``total directional connectedness from others to $j$".

Another important systemic financial risk measure is  $CoVaR$, the value at risk ($VaR$) of an institution $j$, or of the entire the financial system, conditional on an individual institution $i$ being in distress; see \cite{AB2016}.  The $p$-percent $CoVaR$ from firm $i$ to firm  $j$ is defined by 
$$ {CoVaR_{}^{p,j|i}}{:} ~p = P_{}\left( r_{j}<-CoVaR_{}^{p,j|i}~|~ \mathbb{C}
\left( r_{i}\right) \right),
$$
and the $p$-percent $CoVaR$ from firm $i$ to the market is defined by
$$ {CoVaR_{}^{p,mkt|i}}{:} ~
p = P_{}\left( r_{mkt}<-CoVaR_{}^{p,mkt|i}~|~ \mathbb{C}
\left( r_{i}\right) \right).
$$
The leading choice of $\mathbb{C} 	\left( r_{i}\right) $ is a VaR breach, in which case $CoVaR$  measures $VaR$ linkages (hence its name).  The $CoVaR$ directionality, however, is the opposite of $MES$.  It is clearly and intimately related to Diebold-Yilmaz ``total directional connectedness to others from $i$".

\section{Applications}

Many applications followed from the  methodology developed in the above-referenced papers.\footnote{Software implementations that facilitate applications include,  among others, the EViews add-in ``Diebold-Yilmaz Index"; RATS programs to replicate \cite{DieboldandYilmaz2009} (\url{https://estima.com/ratshelp/index.html?dieboldyilmazspilloverpapers.html}) and \cite{DieboldandYilmaz2012} (\url{https://estima.com/ratshelp/index.html?dieboldyilmazijf2012.html}); and R code at \url{https://rdrr.io/cran/Spillover/} and \url{https://github.com/tomaskrehlik/frequencyConnectedness}.  There is also a website, \url{www.financialconnnectedness.org}, with regularly updated dynamic connectedness graphs for major global stock, foreign exchange, sovereign bond, and CDS markets.}  Some were implemented by us.  For example, in \cite{DieboldandYilmaz2009} we obtained time-varying ``total spillover index" (total system-wide connectedness in our current jargon) and showed how global stock return and volatility spillovers behave during major financial crises, such as the East Asian crisis of 1997.  In \cite{DieboldandYilmaz2012} we extended our approach to study directional spillovers and applied the new framework to four financial markets, namely, the U.S. stock, bond, and foreign exchange markets, and the global commodity market.   Going from low-dimensional to high-dimensional environments, \cite{DieboldandYilmaz2014}, \cite{DieboldandYilmaz2016}, and \cite{Demireretal2018} studied connectedness of financial institutions within a country (the U.S.), across the Atlantic, and across the globe, respectively. These  papers put our framework at the center of the literature studying systemic risk in the banking industry, and indeed it has been widely used in policy and industry to assess systemic risk in financial markets.\footnote{Policy examples include IMF Global Financial Stability Reports (\url{https://www.imf.org/en/Publications/GFSR}) and European Central Bank Financial Stability Reviews (\url{https://www.ecb.europa.eu/pub/financial-stability/fsr/html/index.en.html}).}

\section{Concluding Remarks}

With hindsight it is easy to understand the popularity of Diebold-Yilmaz connectedness measurement.  The methodology is simple and appealing,  bridging traditional ``econometric modeling thinking"  on the one hand, and modern ``network and Big Data thinking" on the other, assembling the pieces to go to very new places.  It is based on variance decompositions, which are familiar and comfortable, and it rests on a novel connection between  the seemingly-distinct VAR variance-decomposition literature and the network literature, namely the insight that ``a variance decomposition is a network".  It follows that network tools, which scale effortlessly to high dimensions, provide powerful help in summarizing and visualizing connectedness as defined by variance decompositions.


Interesting recently-published work pushing the frontier outward includes  \cite{barunik2018}, \cite{barigozzi2019},  and 
\cite{bykhovskaya2021}.  Some new unpublished work, moreover, is particularly novel. In one new development, \cite{barigozzi2022} develop methods for connectedness in dynamic ``multilayer" networks, reflecting the insight that rich networks may have many kinds of connections, each governed by its own adjacency matrix.  Parsimonious modeling then becomes absolutely crucial, effectively motivating   \cite{barigozzi2022} to propose a modeling framework with a ``factor structure" for the set of adjacency matrices.  

In a second new development,  \cite{Mlikota2023}  explores  the flip-side of Diebold-Yilmaz thinking.  In particular, while Diebold and Yilmaz  ``turn VARs into networks", Mlikota ``turns  networks into VARs".  More precisely,  Diebold and Yilmaz  map  approximating model variance decompositions into weighted, directed, time-varying networks, which they use to understand dynamic connectedness, whereas Mlikota goes in the opposite direction, modeling approximating model conditional mean functions but exploiting the restrictions implied by network structure to achieve regularization.  Those restrictions take the form of chains, which  generate rich patterns of  multi-step causality \citep{DufourandRenault1998}.

\bibliographystyle{Diebold}
\bibliography{Diebold_DDLY}

%

\end{document}